\documentclass[11pt]{article}
\usepackage{moriond,epsfig}

\def\Journal#1#2#3#4{{#1} {\bf #2}, #3 (#4)}

\def\PLB{{\em Phys. Lett.}  B}
\def\PRL{\em Phys. Rev. Lett.}
\def\PRD{{\em Phys. Rev.} D}

\def\MNRAS{{\em M.N.R.A.S.}}

\def\al{\alpha}

\def\be{\begin{equation}}
\def\ee{\end{equation}}
\def\bea{\begin{eqnarray}}
\def\eea{\end{eqnarray}}

\begin{document}
\vspace*{4cm}
\title{VARYING ALPHA IN THE EARLY UNIVERSE: CMB, LSS AND FMA}

\author{ C. J. A. P. MARTINS }

\address{Centro de Astrof\'{\i}sica da Universidade do Porto, R. das
Estrelas s/n, 4150-762 Porto, Portugal\\
DAMTP, CMS, University of Cambridge,
Wilberforce Road, Cambridge CB3 0WA, United Kingdom \\
and Institut d'Astrophysique de Paris, 98 bis Boulevard Arago,
75014 Paris, France}

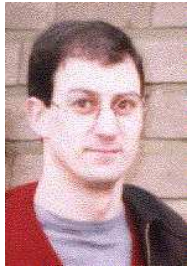
\maketitle\abstracts{
I review recent constraints on and claimed detections of a time-varying
fine-structure constant $\alpha$, and discuss our own work, particularly
on the effects of $\alpha$ in the cosmic microwave background and
large-scale structure.
Our results are consistent with no variation in $\alpha$ from the
epoch of recombination to the present day, and restrict any such
variation to be less than about $4\%$.
The forthcoming MAP and Planck experiments will
be able to break most of the currently existing degeneracies between $\alpha$
and other parameters, and measure $\alpha$ to better than percent accuracy.}

\section{Introduction}
The search for observational evidence for variations of
the `fundamental' constants that can be measured in our
four-dimensional world is an extremely exciting area of
current research, with several independent claims of detections in different
contexts emerging in the past year or so, together with other
improved constraints.
Most of the current efforts have been concentrating on the
fine-structure constant, $\alpha$, both due to its obviously
fundamental role and due to the availability of a series of
independent methods of measurement. Noteworthy among these is
the CMB \cite{Avelino:2000oo,Avelino:2000ea,Battye:2000ds,Avelino:2001nr}.
The latest available CMB results \cite{Avelino:2001nr}
yield a one-sigma indication of a smaller $\alpha$ in the past,
but are consistent with no variation at the two-sigma level.
However, these results are somewhat weakened by the existence of
various important degeneracies in the data, and furthermore not
everybody agrees on which and how strong these degeneracies
are \cite{Avelino:2000ea,Battye:2000ds,Huey:2001ku}.

We \cite{us} have recently
obtained up-to-date contraints on variations of
$\alpha$ from CMB and large-scale structure data, as
well as analyzed these
possible degeneracies in some detail, mainly by means of a
Fisher Matrix Analysis (FMA). There
are crucial differences between `theoretical' degeneracies
(due to simple physical mechanisms) and `experimental'
degeneracies (due to the fact that each CMB experiment
only probes a limited range of scales, and that the experimental
errors are scale-dependent).
Such degeneracies can of course be eliminated either by using
complementary data sets (such as large-scale structure constraints)
or by acquiring better data (such as that to
be obtained by MAP and Planck). Finally, a further, more elegant
way of doing this is by including
information from CMB polarization \cite{Rocha:2002}.

\section{The Current Observational Status}

The recent explosion of interest in the study of varying constants is
mostly due to the results of Webb and
collaborators \cite{Murphy:2000pz,Webb:2000mn,Murphy:2000ns,Murphy:2001nu}
of a $4\sigma$ detection of a fine-structure constant that was smaller
in the past, ${\Delta\alpha}/{\alpha} = (-0.72 \pm 0.18)\times
10^{-5}$ for $z\sim0.5-3.5$;
indeed, more recent work \cite{Webb:2001} provides an even stronger
detection.
An independent, simpler technique was used to measure the ratio of
the proton and electron masses, $\mu=m_p/m_e$ \cite{Ivanchik:2001ji}.
Using two systems at
redshifts $z\sim2.3$ and $z\sim3.0$ are
${\Delta\mu}/{\mu} = (5.7 \pm 3.8)\times 10^{-5}$
or
${\Delta\mu}/{\mu} = (12.5 \pm 4.5)\times 10^{-5}$
depending on which of the (two) available tables of `standard'
laboratory wavelengths is used. This implies a $1.5 \sigma$
detection in the more conservative case, though it also casts some
doubts on the accuracy of the laboratory results, and on the
influence of systematic effects in general.

A recent re-analysis \cite{Fujii:2002bi} of the Oklo bound using new
Samarium samples
collected deeper underground (aiming to minimize contamination)
again finds two possible results,
${\dot\alpha}/{\alpha}=(0.4 \pm 0.5)\times 10^{-17} yr^{-1}$
or ${\dot\alpha}/{\alpha}=-(4.4 \pm 0.4)\times 10^{-17} yr^{-1}$
Note that these are given as rates of variation, and effectively
probe timescales corresponding to a cosmological redshift of about
$z\sim 0.1$. Unlike the case above, these two values correspond
to two possible physical branches of the solution.
(Note that these results have opposite signs to previously published ones).
The first of branch provides a null result, while
the second is a strong detection of an $\alpha$ that was
{\it larger} at $z\sim0.1$, that is a relative variation
that is opposite to Webb's result. Even though there
are some hints (coming from the analysis of Gadolinium samples)
that the first branch is preferred, but this is by no means settled
and further analysis is required to verify it.

One can speculate about the possibility that the second branch
is the correct one---this would definitely
be the most exciting possibility. While in itself this wouldn't
contradict Webb's results (since Oklo probes much smaller
redshift and the suggested magnitude of the variation is smaller
than that suggested by the quasar data), it would have
striking effects on the theoretical modelling of such variations.
Proof that $\alpha$ was once larger
would sound the death knell for any theory which models
the varying $\alpha$ through a scalar field whose behaviour
is akin to that of a dilaton. Examples include Bekenstein's theory
\cite{Bekenstein:1982eu} or simple variations thereof
\cite{Sandvik:2001rv,Olive:2001vz}.
Indeed, one can quite easily see
\cite{Damour:1993id,Santiago:1998ae} that in any such model
having sensible cosmological parameters and obeying other
standard constraints 
$\alpha$ must be a monotonically increasing function of time.
Since these dilatonic-type models are arguably the simplest
and best-motivated models for varying alpha from a particle
physics point of view, any evidence against them would be
extremely exciting, since it would point towards the presence
of significantly different, yet undiscovered physical mechanisms.

Finally, we also mention that there have been recent proposals
\cite{Braxmaier:2001ph} of more accurate laboratory tests of
the time independence of $\alpha$ and the ratio of the
proton and electron masses $\mu$, which could improve current
local bounds by an order of magnitude or more.

However, given that there are both theoretical and experimental
reasons to expect that any recent variations will be small, it
is crucial to develop tools allowing us to measure $\alpha$ in
the early universe, as variations with respect to the present value
could be much larger then.
In our previous
work \cite{Avelino:2001nr}, we have carried out a joint
analysis using the most recent CMB (BOOMERanG and DASI) and
big-bang nucleosynthesis (BBN) data, finding evidence at the one sigma
level for a smaller alpha in the past (at the level of $10^{-2}$ or
$10^{-3}$), though at the two sigma level the results were
consistent with no variation. However, as can be seen by comparing
with earlier work \cite{Avelino:2000ea,Battye:2000ds} (and has
also been discussed explicitly in these papers), these
results are quite strongly dependent on both the observational
datasets and the priors one uses.
This begs the question of whether
the degeneracies found in \cite{Avelino:2000ea,Battye:2000ds,Avelino:2001nr}
are real `physical' and fundamental degeneracies, which will remain at
some level, no matter how much more accurate data one can get, or
if they are simply degeneracies in the data, which won't
necessarily be there in other (better) datasets. And a related
question is, of course, assuming that the degeneracies are
significant, how can one get around them.

\section{Current CMB and Large-Scale Structure Constraints}

We have recently performed \cite{us} an up-to-date analysis of the
Cosmic Microwave Background
constraints on varying $\alpha$ as well as, for the first time,
an analysis of its
effects on the large-scale structure (LSS) power spectrum.
See \cite{us} for a discussion of why both of these are
affected by variations in $\alpha$.

Although these CMB and LSS constraints are
complementary, and can help break degeneracies
by determining other cosmological parameters, they certainly
can not be blindly combined together.
We are not combining direct constraints
on the parameter $\alpha$ itself, obtained through both methods, to
obtain a tighter constraint. This can not be done,
since the CMB and LSS analyses are sensitive to the values of $\alpha$
at different redshift ranges, so there is no reason why these values
should be the same. Additionally, there is no well-motivated theory that
could relate such variations at different cosmological epochs. All that
one could do at this stage would be to assume some toy model where a
certain behaviour would occur, but this would mean introducing various
additional parameters, thus weakening the analysis. We chose not
to pursue this path and leave the analysis as model-independent as possible.
What we are doing is using additional information (which is also
sensitive to $\alpha$) to better constrain other parameters in the
underlying cosmological model, such as $n_s$, $h$ and the densities of
various matter components, which we can reliably assume are
unchanged throughout the cosmological epochs in question. In other words,
we are simply self-consistently selecting more stringent priors
for our analysis.

The constraints obtained by combined analysis are reported
in \cite{us}. As one would expect,
when constraints from 
other and independent cosmological datasets are included in the
CMB analysis, the constraints on variations on $\alpha$ become
significantly stronger.

\section{Fisher Matrix Analysis}

The precision with which the forthcoming satellite 
experiments MAP and Planck will be able to
determine variations in $\al$ can be readily 
estimated with a Fisher Matrix Analysis (FMA).
Some authors have already performed such an analysis in the past, 
but their analysis was based on a 
different set of cosmological parameters and assumed cosmic variance 
limited measurements. In our FMA \cite{us} we also take into account 
the expected performance of the MAP and Planck satellites and we make use of a 
cosmological parameters set which is well adapted for limiting numerical 
inaccuracies. Furthermore, the FMA can provide useful insight into 
the degeneracies among different parameters, with minimal computational effort.

MAP will be able to constrain variations in $\alpha$ at 
the time of last scattering to
within $2\%$ ($1\sigma$, all others marginalised). This
corresponds to an improvement of a factor of 3 relative to current limits,
Planck will narrow it down
to about half a percent. If all other parameters are supposed to be known
and fixed to their ML value, then a factor of 10 is to be gained in the
accuracy of $\al$. However, if all parameters are being estimated
jointly, the accuracy on variations in $\alpha$ will not go beyond
$1\%$, even for Planck.
The parameters $\omega_b, {\mathcal R}$ and $n_s$ suffer
from partial degeneracies with $\al$, which are discussed in more detail
in \cite{us}.
Planck's errors, as measured by the inverse square root of the
eigenvalues, are smaller by a factor of about 4 on average. For all but one
eigenvector Planck also obtains a better alignment of the principal
directions with the axis of the physical parameters.
This is of course in a slightly different
form the statement that Planck will measure the cosmological parameters
with less correlations among them.

\section{Conclusions}

We have shown that the currently available data is consistent with no
variation of $\alpha$ from the epoch of recombination to the
present day, though interestingly enough the CMB and LSS
datasets seem to prefer, on their own, variations of $\alpha$ with
opposite signs. Whether or not this statement has any physical
relevance (beyond the results of the statistical analysis) is
something that remains to be investigated in more detail. In any case,
any such (relative) variation is constrained to be less than
about $4\%$.

The prospects for the future are definitely bright.
In the short term, the VSA and CBI data (first announced 
at Moriond) will provide some improvement
on the current results.
In the longer term, the forthcoming satellite experiments will provide
a dramatic improvement on these results.
These tools, together with other
measurements coming from BBN \cite{Avelino:2000ea} and quasar and
related data \cite{Murphy:2000pz,Webb:2000mn} offer the exciting prospect
of being able to map the value of $\alpha$ at very many different cosmological
epochs, which would allow us to impose very tight constraints on
higher-dimensional models where these variations are ubiquitous.

\section*{Acknowledgments}
It is a pleasure to thank Pedro Avelino, Rachel Bean, Alessandro
Melchiorri, Gra\c ca Rocha, Roberto Trotta and Pedro Viana for a
very enjoyable collaboration.
This work is partially supported by the European Network CMBNET,
and was performed on COSMOS, the Origin2000 owned by the UK
Computational Cosmology Consortium, supported by Silicon Graphics/Cray
Research, HEFCE and PPARC.
C.M. is funded by FCT (Portugal), under grant no. FMRH/BPD/1600/2000.

\end{document}